\documentclass[12pt]{article}

\usepackage{setspace} 

\usepackage{authblk}
\usepackage{natbib}
\usepackage{graphicx}
\usepackage[colorlinks=true, allcolors=blue]{hyperref}

\def\spacingset#1{\renewcommand{\baselinestretch}%
{#1}\small\normalsize} \spacingset{1}

\begin{document}

\title{Open-source Tools for Training Resources - OTTR}

\author[1]{Savonen, Candace}
\author[1]{Wright, Carrie}
\author[1]{Hoffman, Ava M.}
\author[1]{Muschelli, John}
\author[1]{Cox, Katherine}
\author[2]{Tan, Frederick J.}
\author[1]{Leek, Jeffrey T.}

\affil[1]{Johns Hopkins Bloomberg School of Public Health, Baltimore, MD}
\affil[2]{Carnegie Institution, Baltimore, MD}
\affil[*]{Corresponding author: \href{mailto:csavone1@jhu.edu}{csavone1@jhu.edu}}

\maketitle

\begin{abstract}
Data science and informatics tools are developing at a blistering rate, but their users often lack the educational background or resources to efficiently apply the methods to their research. 
Training resources often deprecate because their maintenance is not prioritized by funding, giving teams little time to devote to such endeavors. 
Our group has developed \href{https://github.com/jhudsl/OTTR_Template/wiki}{Open-source Tools for Training Resources (OTTR)} to offer greater efficiency and flexibility for creating and maintaining online course content.
OTTR empowers creators to customize their work and allows for a simple workflow to publish using multiple platforms.
OTTR allows content creators to publish material to multiple massive online learner communities using familiar rendering mechanics. 
OTTR allows the incorporation of pedagogical practices like formative and summative assessments in the form of multiple choice questions and fill in the blank problems that are automatically graded. No local installation of any software is required to begin creating content with OTTR.
Thus far, 15 courses have been created with OTTR repository template. By using the OTTR system, the maintenance workload for updating these courses across platforms has been drastically reduced.

\end {abstract}

\noindent%
\vfill 

\newpage
\spacingset{1.45}
\section{Introduction}

Informatics, data science, and machine learning tools are developing at a dramatic rate. Driven by the rapidly changing technologies that underlie informatics and data science, such as genomic sequencers and imaging technologies, tool developers are constantly working toward new software and updates to handle these data. However, their users often lack the educational background or resources to efficiently apply these quickly evolving methods to their research \cite{Attwood2017, Wren2016}. 
Training resources and documentation for these tools are too often non-existent or left to deprecate \cite{Mangul2019}. 
Those with the deepest and most up-to-date knowledge of the tools, such as developers and power users, generally have limited time or resources to devote to training efforts \cite{Kaiser2016}.

Unfortunately, a large portion of time spent on training material creation is not related to content and does not require content-specific knowledge. Time is instead spent determining where and how to publish and maintain the material as opposed to what the material should contain. These endeavors are especially time consuming for individuals if they have no training in creating online educational content. Overall, this creates a frustrating and inefficient dynamic for both developers and their end users.

Training communities such as \href{https://software-carpentry.org/}{The Carpentries} and \href{https://training.galaxyproject.org/}{Galaxy Training Network} have led the way in creating community resources to improve the development and maintainability of informatics and data science education \cite{Wilson2016, Batut2018}. These organizations have their own sets of tools, guidelines and templates for individuals who would like to contribute content to their community platform (See \hyperref[fig:compareplatforms]{Supplementary Table 1}). We have created similar tools and templates that allow for large-scale, flexible, and customize-able publishing in a simplified way as they are... 
\begin{itemize}
    \item Compatible with massive open online course platforms like \href{https://www.coursera.org/}{Coursera} and \href{https://leanpub.com/}{Leanpub} which have millions of users \cite{Coursera2021, Leanpub2021}
    \item Designed to encourage customization of branding and publishing platform options.
    \item Built for easy maintenance to encourage the upkeep of the material for a longer period of time. 
\end{itemize}

We have developed the \href{https://github.com/jhudsl/OTTR_Template/wiki}{Open-source Tools for Training Resources (OTTR)} to offer the efficiency of templated setups that The Carpentries and Galaxy Training Network have, but with a focus on creating customizable Massive Open Online Courses (MOOCs). 
OTTR is a system of tools, templates, and guides that supports course creation in addition to publication to multiple massive online learner communities that already exist on Coursera and Leanpub. 
OTTR's compatibility with Coursera and Leanpub provides a method for incorporating pedagogical practices like formative and summative assessments in the form of multiple choice questions and fill in the blank problems that are automatically graded. MOOCs like Coursera and Leanpub help democratize education by allowing learners to acquire certification for completing courses in a way that is well-recognized and can help learners improve their employment opportunities \cite{Rivas2020}. 
OTTR also enables customization of style and course accessories so that courses can be tailored for their context, including both the course's specific audience and content creator's institution. Importantly, our tools offer developers the opportunity to write content once and publish in multiple ways if they choose. Our tools and template also allow users flexibility in how they would like their content published. Users can pick between publishing methods and strategies, including what platforms to publish on and if they would like to include quizzes or not.

Anyone can create material through OTTR at any time without prior approval or additional accounts; the only requirement is a free GitHub account, and only git is needed; a git client like \href{https://www.gitkraken.com/}{GitKraken} is strongly recommended for those unfamiliar with GitHub \cite{github2022, gitkraken2022}. We believe OTTR builds upon existing models of content creation to strengthen the dynamic between developers and users, while also improving the democratization of informatics and data science-related education \cite{Kross2020a}.

\subsection{OTTR Design Philosophy}

\begin{itemize}
\item \textbf{Maintainable} - After creating training content, informatics tool developers and content creators often move on to other projects or have other priorities take the forefront. 
We aim to make the workload for future course updates and maintenance as light as possible to deter deprecation of the training resources. 
Building as much as possible programmatically from plain text (e.g. R Markdown) using \href{https://www.docker.com/}{Docker} images produces the same content every time while ensuring proper environment and version control \cite{Merkel2014}. 
Easy maintenance also allows developers and content creators to focus on the cutting edge of their fields, rather than time-consuming upkeep of training materials across platforms. 
The pull request model of GitHub also allows contributions from the community, which could reduce the burden on the original tool developers and increase the robustness of the material. OTTR's automatic checks can help polish any incoming contributions, helping alleviate the cognitive load. 
\item \textbf{Adaptable} - Training resources might be needed in many different contexts. We aim to be prescriptive but flexible.  Permissive licensing like CC-BY allows content creators to quickly mix and match snippets of code or subsets of slides and customize for a specific audience. All automated tools are customize-able from a simple central configuration file that a user can set on and off tools depending on their course's needs. From this file, users can turn off or on the renderings and checks for publishing to one platform or the other if they do not want to publish to all three platforms. This means the user can avoid the risk of being entirely committed to the file formats required by a particular platform as they are creating content. Furthermore, course chapters that are useful in multiple courses, \href{https://github.com/jhudsl/OTTR_Template/wiki/Borrowing-chapters-between-courses}{can be borrowed from one origin} minimizing redundant maintenance. 
\item \textbf{Open source and free} - Funding is not always available for developing or maintaining training material; we do not want that to be a barrier to use these tools. A lack of funding for developing materials also propagates existing financial inequities among under-served institutions and faculty \cite{Hemming2019}. Ensuring tools like OTTR do not require the purchase of any software can help alleviate these challenges. Using GitHub's free-tier hosting and workflow means its inherently open source and free. 
\item \textbf{Templates for momentum} - Creating material from a “blank slate” is challenging, and decision fatigue can be a hindrance to authoring new training resources. We aim to give course creators sensible defaults and templates so these creators can begin writing content as soon as possible. But we also provide functionality and guidance for users to customize their material as they need, though use of custom style sets and Docker images.
\end{itemize}

\spacingset{1.45}
\section{Methods}

\begin{figure}
\centering
\includegraphics[width=.8\linewidth]{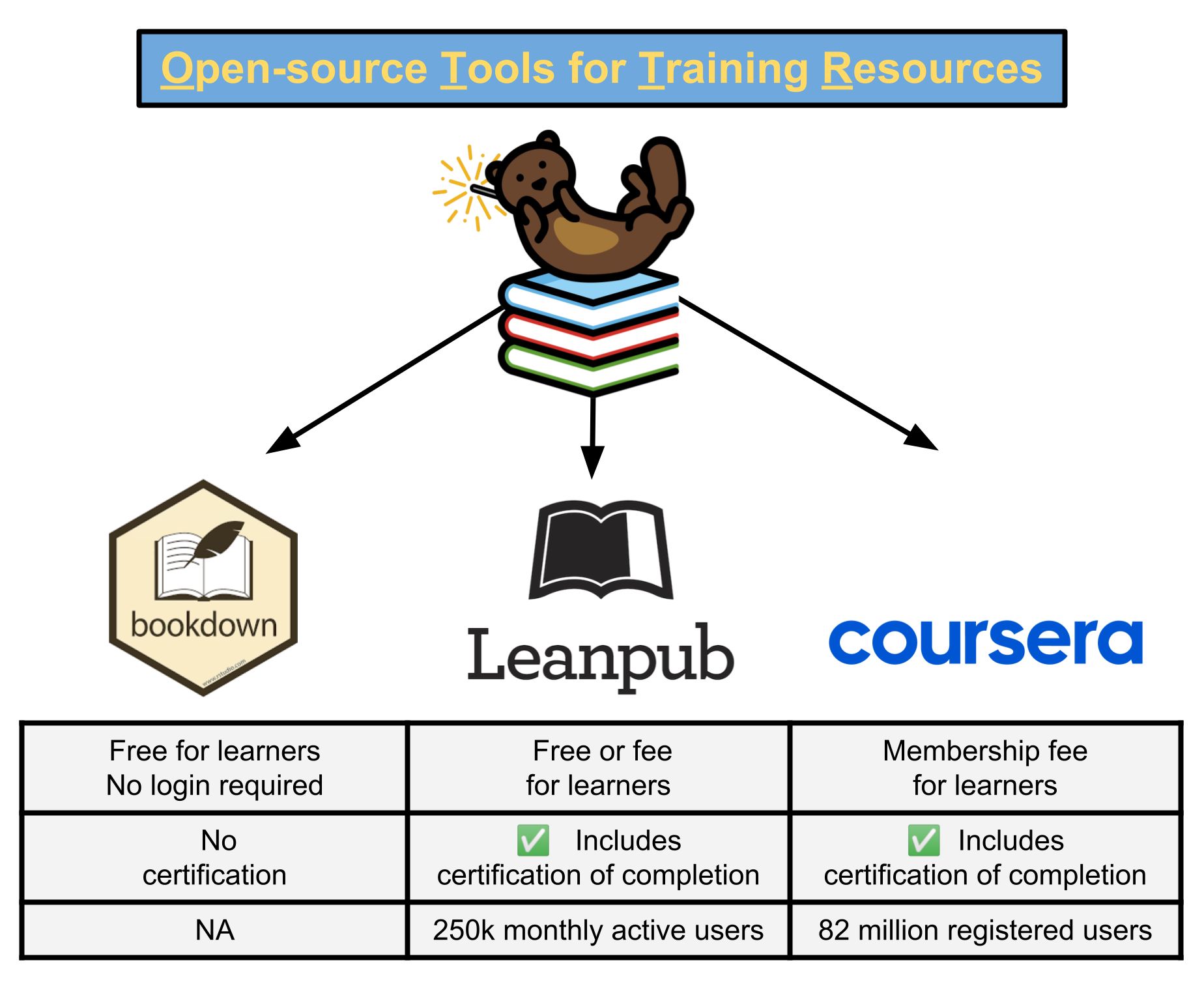}
\caption{Publishing platform options. OTTR can publish to three platforms which each have distinct audiences and meet different learner needs. The OTTR system minimizes the maintenance pains for upkeep on these three platforms by allowing you to write once and have updates to the material automatically sent to all platforms.}
\label{fig:summary}
\end{figure}

With OTTR, we have devised a system that allows content creators to generally write once, but publish in three places.  Bookdown is an R package and system to create content using RMarkdown, which is a text-based format where you can embed code, plots, math equations, references, and a number of other features of word processing.  This RMarkdown content can be rendered as a PDF, ePub, or HTML page, so this fits with a number of readers, namely e-readers and website hosting. 
A Bookdown website is the default OTTR platform and is hosted with GitHub pages upon set up. 
Leanpub and Coursera allow the distribution of this content, either as a book (Leanpub) or a course (Leanpub and Coursera).  
Publishing with Bookdown, Leanpub, and Coursera not only allows the course content to reach a wider audience but also allows learners to pick the learning platform that best suits their needs or preferences \cite{Xie2016, Leanpub2021, Coursera2021}. The Bookdown-based website is similar to a course website that organizes content in a logical way via chapters or modules. Some content creators might find Bookdown to be the most approachable of the three publishing platforms because it only requires one step for user access: navigating to the appropriate URL. Because the Bookdown-rendered website does not require any login or user account, it cannot provide any certificate of completion. In contrast, Leanpub and Coursera platforms offer quizzes and certificates of completion. This allows user audiences to better assess and keep a record of their learning. Leanpub allows content authors to set their own price for end users (even free) while Coursera requires a paid subscription (See Figure~\ref{fig:summary}).

Publishing in all three platforms has many benefits but can quickly become a maintenance burden. 
OTTR uses automation and focuses on features that can be well-supported across MOOC platforms to minimize these maintenance pains. It also provides built-in flexibility for creators to change or add publishing platforms at any time during the content creation process, minimizing the risk associated with "committing" to a specific platform.

\subsection{The Writing Process}

The center of the OTTR approach is writing an R Markdown document \cite{rmarkdown2022}. 
R Markdown extends the \href{https://www.markdownguide.org/}{Markdown} specification with some special capabilities based in R. Generally speaking, R Markdown allows the author to combine text, images, and code in one plain text document prior to rendering. This allows for flexibility of content; courses are written in traditional Markdown language but can also render example code not only for R but has extension capabilities for other languages (like Python).

\textbf{Prerequisite knowledge for OTTR usage:}
\begin{itemize}
\item Familiarity with GitHub.
\item Familiarity with Markdown.
\item Very basic R knowledge is helpful but not required (local R installation is not needed). 
\end{itemize}

For individuals who do not have this prerequisite knowledge but still wish to use OTTR, \href{https://github.com/jhudsl/OTTR_Template/wiki/Getting-started#recommended-background-information}{the OTTR Guide} points authors to resources to gain familiarity with these tools. 

\subsection{Rendering and Previewing}
Rendering (and re-rendering) is automated by \href{https://github.com/features/actions}{GitHub Actions}, which is a service that automatically runs a set of directions or scripts whenever a change is made, meaning that course creators can create content with their preferred text editor. 
This eliminates the need for local installation of any additional software. Focusing on just the R Markdown files (and not the resulting rendered files) also prevents confusion that may be introduced when tracking binary files from multiple authors' file changes. Content can be edited directly from the GitHub web editor and pull requests created from the GitHub website. 

When content is pushed to an open pull request on an OTTR-templated GitHub repository, it is rendered automatically as HTML using Bookdown on a \href{https://pages.github.com/}{GitHub Pages} website (See Figure~\ref{fig:preview}) \cite{Xie2016}. This rendering includes an option to include other outputs, such as .pdf and .docx files which allow easy sharing with collaborators. Docx files can be uploaded to Google Drive for easy collection of feedback from collaborators.
A preview of the latest content and changes are automatically provided in a GitHub comment (See Figure~\ref{fig:preview}).

\subsection{Publishing}
\begin{figure}
\centering
\includegraphics[width=1\linewidth]{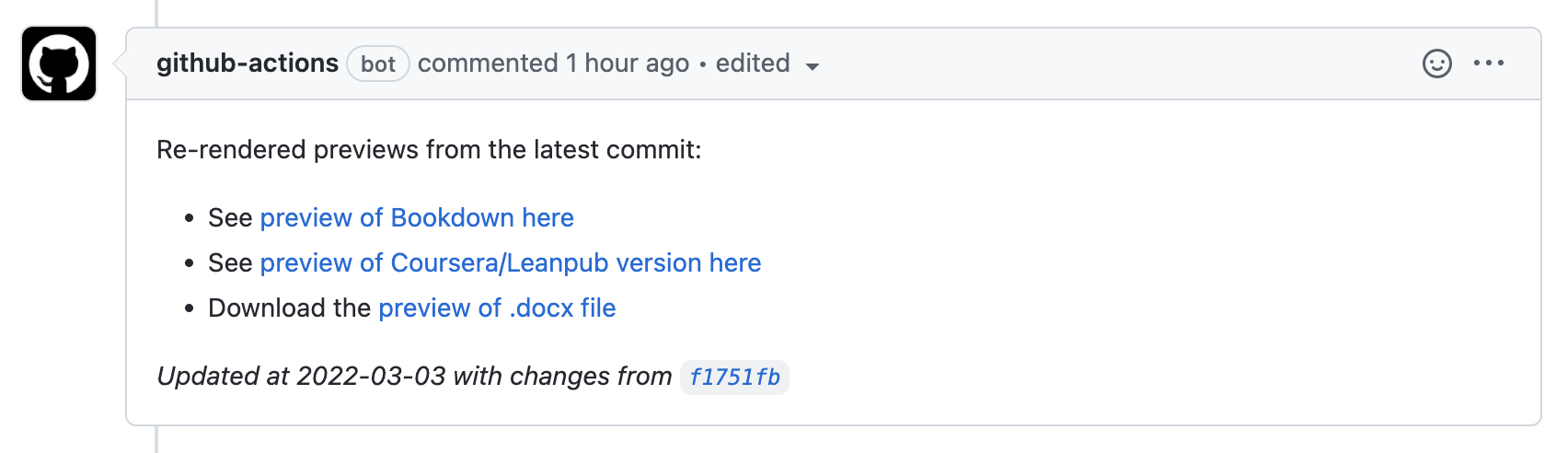}
\caption{Automatic rendering. Upon opening a pull request, OTTR creates a preview of the changes in different formats so the content creator can easily review the changes. The preview is provided in three ways: A link to what the Bookdown rendered website would look like, a link to what the Coursera/Leanpub material iframe preview would look like, and a docx file which allows for use of track changes with Google Docs or Microsoft Word. }
\label{fig:preview}
\end{figure}

The automatically rendered Bookdown version of the course is published using GitHub Pages. 
Optionally, a course creator can connect individual chapters on Leanpub and Coursera following the \href{https://github.com/jhudsl/OTTR_Template/wiki/Choosing-publishing-platform(s)}{OTTR Guide}. 
A custom domain can host a published GitHub Pages so the content creator may choose where to host this material.  
Once this set up is done, updates made within a chapter are automatically available in Coursera and Leanpub as soon as the changes are made live on the Bookdown course (See Figure~\ref{fig:preview}).

\begin{figure}
\centering
\includegraphics[width=1\linewidth]{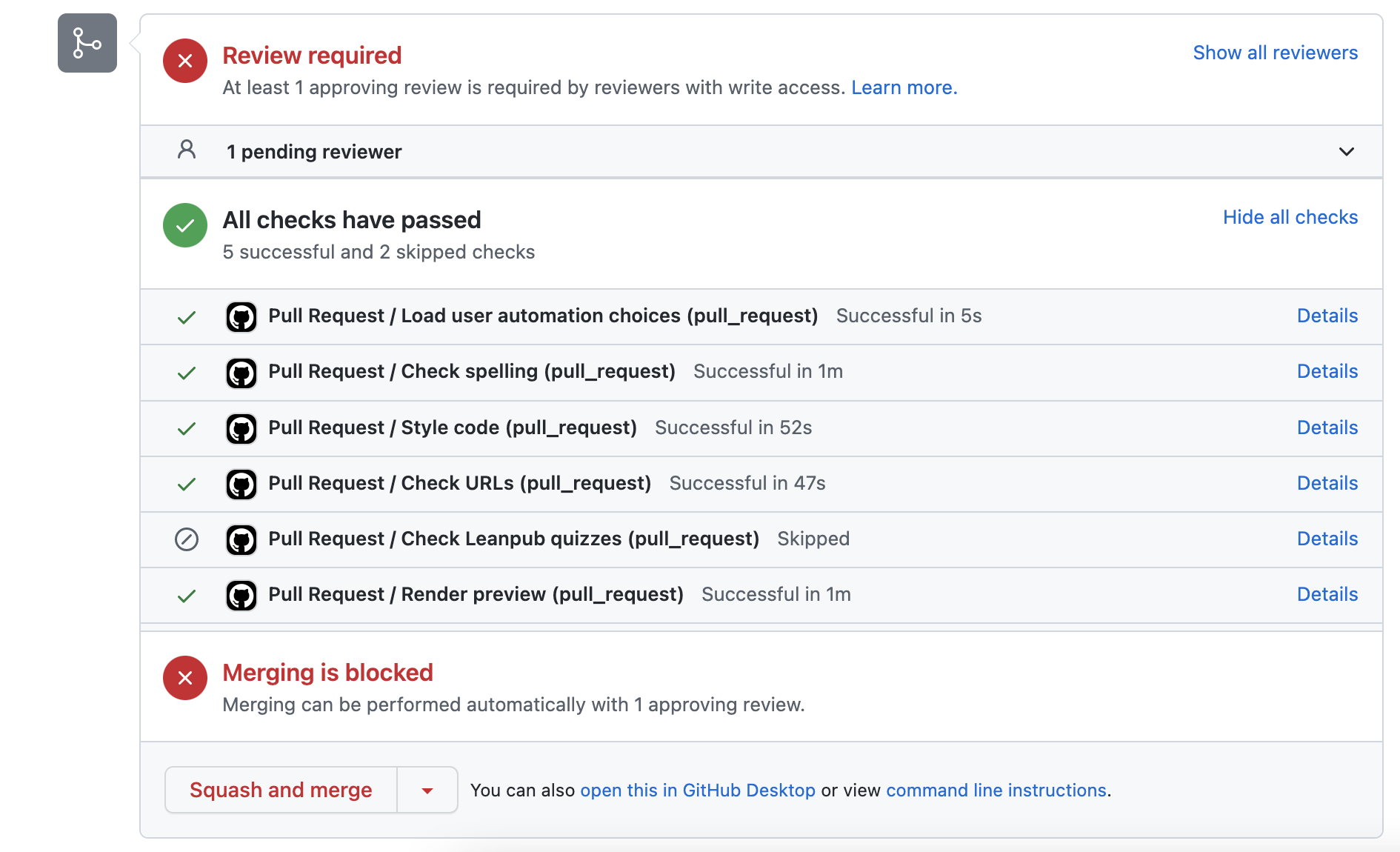}
\caption{Automatic checks. Upon opening a pull request, OTTR runs a series of customize-able checks on the new content. These can be turned on or off by a central configuration file. }
\label{fig:checks}
\end{figure}

\section{OTTR Components}
As our team developed our own courses, we looked for opportunities in the process for minimizing the workload of ourselves and other course creators. 
For each time consuming process identified, we attempt to create a technical solution such as: an automated tool, a template, or a guide to ease the process. 

\subsection{Automated tools}
Not everything can or should be automated, but the more menial the task, the more that automation is not only possible but is preferred to manually performing the task.
Manually performing menial tasks not only is an inefficient use of a course creator's time, but automated processes are generally more thorough at completing and checking certain tasks. For example, HTML pages are automatically rendered by GitHub Actions. This saves the course creator's time, avoids accidentally missing files to be rendered, and prevents conflicts when the same content has been edited in multiple ways.

Previews and checks of new content happens automatically as changes are proposed in the form of an open pull request.
A pull request is a way to highlight prepared changes to the content. This allows others to review and comment on potential changes before they are incorporated into the live content.
The OTTR workflow is centered around the pull request to encourage collaborative review.
To this end, the automatic checks and previews occur when pull requests are opened and changes are added; the results of these checks are posted for OTTR authors to easily see and download on the pull request page.
The outcomes of checks of spelling, quiz formatting, Docker image building, and working URLs are all reported on pull requests (See Figure~\ref{fig:checks}). 
All these automated tools, checks, and renderings can be turned on or off from a central configuration file.

\subsubsection{Reproducibility using Docker}
Docker is a containerization tool which allows content creators to use the same computing environment regardless of their local operating system, essentially freezing the environment to complete rendering tasks so that it does not break with future renderings, especially of the same material.
Docker is advantageous to use in the content creation process because it is lightweight (i.e., minimal storage needs) and ensures rendering consistency across collaborators on a set of content.
OTTR provides and uses a Docker image to render course material automatically using GitHub Actions. 
This Docker image can also be used locally as an image to build upon for other needs, but local Docker installation or knowledge is not a requirement for using OTTR. Along with Git and GitHub, Docker is a key component of reproducible research \cite{Nust2020}. 
Containerization avoids mismatches in R version, package behavior, or hierarchical dependency (otherwise known as ``dependency hell''). 
To maximize reproducibility for creators, we have implemented a \href{https://hub.docker.com/r/jhudsl/course_template}{container specifically for OTTR}, based on the Rocker family of Docker containers \cite{Boettiger2017}. 
Because Docker might be less familiar to creators compared with GitHub or Markdown, our guide also contains further description on \href{https://github.com/jhudsl/OTTR_Template/wiki/Using-Docker}{how to customize and adapt this Docker image} depending on the particular needs of a course.

\subsubsection{\textit{ottrpal} Companion Package}

The \href{https://github.com/jhudsl/ottrpal}{\textit{ottrpal}} R package, a companion package we developed for the OTTR template, assists with the publishing process. One of the functions \textit{include\_slide()} helps users include images from Google slide decks more seamlessly. Using Google Slides for image storage allows for \textit{ottrpal} to update images seamlessly. Users specify alternative text and other code chunk options, but upon re-rendering the images are all automatically updated from the linked Google Slide deck.  Thus, the author does not need to copy and paste updated slides, there is one source of truth for embedded images (in the Google Slide deck), and the content is referenced where it came from.  Google slides also, in their speaker notes, allows the linking of metadata about the slide that can seen as additional documentation for content.

After an initial set up on Leanpub and Coursera, no manual updating is needed for typical updates through the use of embed windows (iframes) on these platforms that link to the main Bookdown content.
This means Coursera and Leanpub are instantly refreshed upon course updates to the course on GitHub. 
\textit{ottrpal} package creates these embed window versions for Coursera and Leanpub automatically. This simplifies maintenance across platforms so chapter updates do not require extra manual steps (Figure~\ref{fig:maintenance}).

The \textit{ottrpal} package also facilitates format conversion and quiz creation for uploading to Leanpub and Coursera. 
It also performs checking of quiz formatting to ease the quiz upload process to Leanpub and converts Leanpub quizzes to a format ready to upload to Coursera.  Without OTTR and \textit{ottrpal}, these formatting changes must be done by authors, which requires knowledge of the specifics of each platform.

\subsection{Templates}
\subsubsection{Quick Start Template Files}
The OTTR repository itself is a \href{https://docs.github.com/en/repositories/creating-and-managing-repositories/creating-a-template-repository}{GitHub template repository}, which means that it can be used to create a new repository but this new repository will have its own fresh history. 
It contains R Markdown templates that demonstrate examples of writing content as well as pull request templates, and issue templates to help facilitate discussions. 
The pull request template also has a check list to help guide the incorporation of new content. 

Upon creation of a new repository based on the OTTR template repository, GitHub issues are automatically filed to help course creators follow along to set up their repository. GitHub issues track the to-do list OTTR-users can follow for the initial set up. 

\subsubsection{Authorship and Credits Template}
Course creation often involves a lot of individuals and the tasks accomplished do not generally fit into the typical authoring system used by academia. Moreover, the existing system for crediting authors in scholarly work has been widely criticized for opacity and lack of accountability and/or deserved credit \cite{Borenstein2015}. 
We have devised a templated system of \href{https://github.com/jhudsl/OTTR_Template/wiki/How-to-give-credits}{credit labels} for course contributors which are reported in a table in the course. 
This credit system is analogous to a film credit system where one individual may have had multiple roles, and roles are shown in an order of highest to lowest level of involvement. We also gained inspiration from the \href{https://casrai.org/credit/}{"CRediT" taxonomy}. 

\subsubsection{Course Feedback Templates}
In order to make sure a course is properly serving its learners, it needs a mechanism for learners to report concerns and general feedback. 
On each platform, we have incorporated a templated system that links to a course creator's chosen method of feedback, such as via GitHub issues or anonymous Google Forms.
This way, no matter how a learner chooses to take the course, there is always a readily available mechanism for providing feedback, and in turn, provide continuous pathways for course improvement. 

\subsection{Detailed Guides}
Where automation or additional tools are not reasonable avenues, we have opted to \href{https://github.com/jhudsl/OTTR_Template/wiki}{document the process and choices}. 
This guide covers the set up of the repositories and the platforms. 
It also provides advice and tips for course creation and available tools. 
Upon usage of the GitHub template and the creation of a new course repository, the automatic issues filed link to this guide in order to help the course creators track and follow along with the guide for set up. 

\subsection{OTTR Version Updates}

As OTTR is updated and new features are added, changes to the mechanics can be incorporated into downstream courses that utilize the OTTR template (updates are an opt-in system at this point in time). 
When OTTR has a new release, this change is offered to downstream repositories that have opted-in for updates.
These changes are offered in the form of a pull request that course authors can feel free to incorporate as they see fit. 
These new OTTR feature pull requests uses the \href{https://github.com/marketplace/actions/repo-file-sync-action}{repository file sync} GitHub action \cite{Schiller2021}.
This allows us to reasonably let course creators take advantage of new OTTR features on their course repositories that predated the initial creation of the course repository. 

\section{Examples}

\begin{figure}
\centering
\includegraphics[width=.8\linewidth]{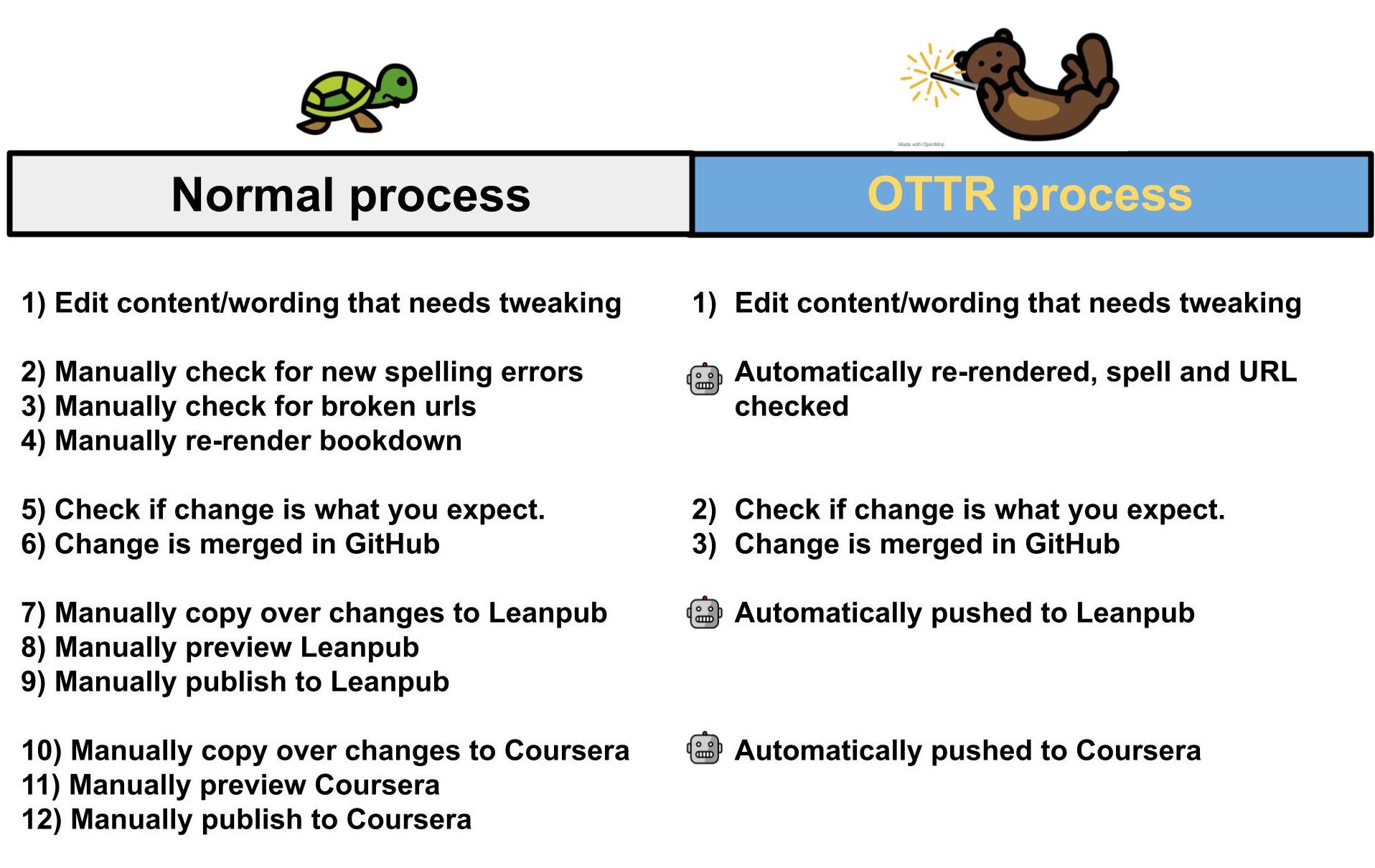}
\caption{Maintenance steps with and without OTTR. Note that the steps with a robot icon are done automatically in the OTTR setup. }
\label{fig:maintenance}
\end{figure}

Thus far, 15 courses across 3 institutions have been created with OTTR repository template. Many of these have resulted in courses published on Leanpub and Coursera.  
By using the OTTR system, the maintenance workload for these updates has been drastically reduced.
Figure~\ref{fig:maintenance} shows a comparison of the steps involved in the case of a typical wording update to a course. 
OTTR simplifies updates, making maintenance and minor fixes to courses much more likely to be completed. In the long run, this should help prolong the relevance of these courses.

\begin{figure}
\centering
\includegraphics[width=.5\linewidth]{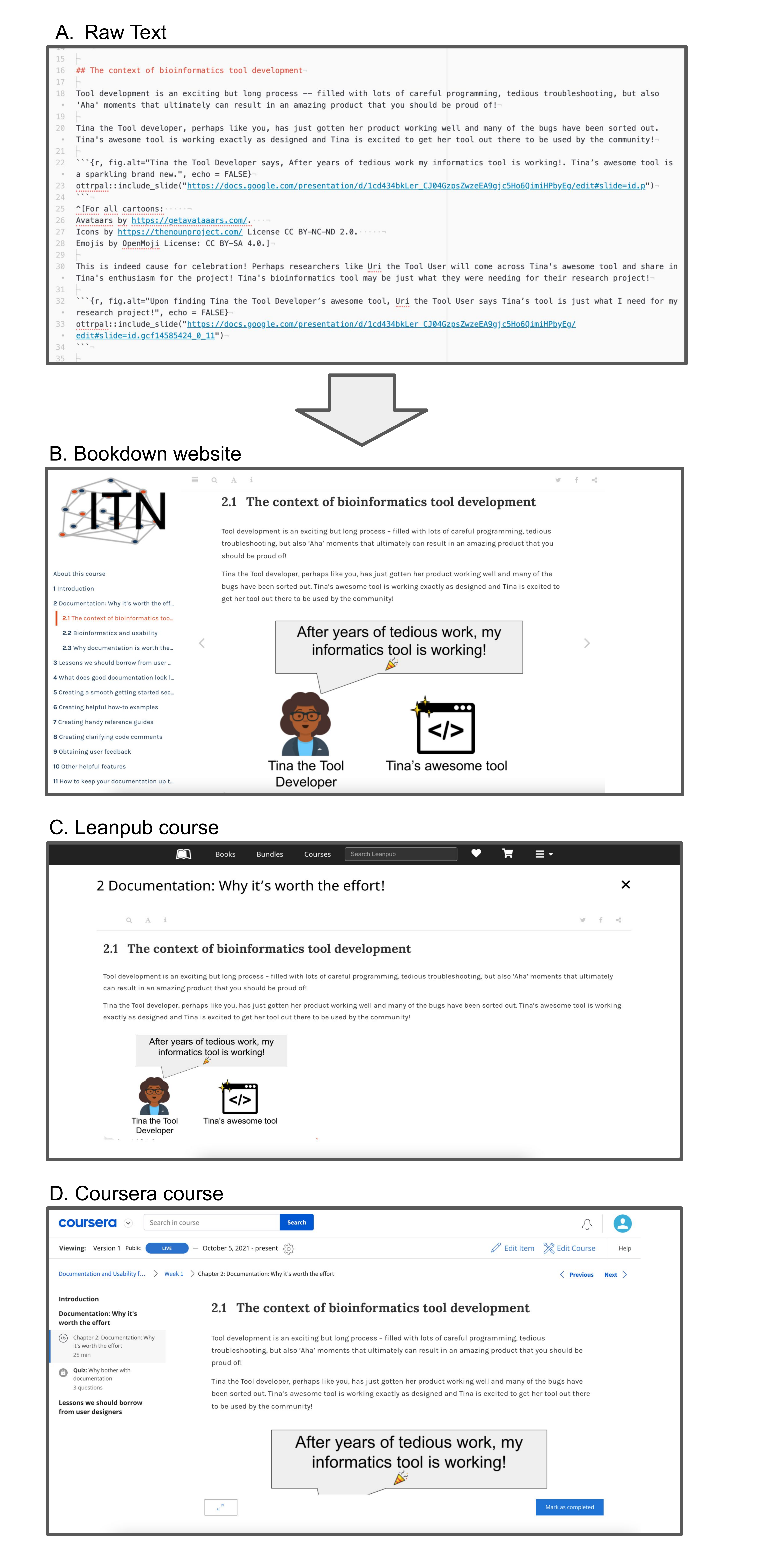}
\caption{Example of rendered content. The first panel (A) shows the raw text for this example course page made with OTTR. It shows how the raw text written by a content author might look like as (B) a Bookdown website hosted by GitHub pages, (C) a Coursera course and (D) Leanpub course. This automatic rendering has an added benefit of keeping the new file changes focused on the content changes instead of html rendering artifacts.}
\label{fig:rendering}
\end{figure}

One such OTTR course regarding Documentation and Usability for Cancer Informatics is shown here as an example \cite{Savonen2021}. 
This course's own \href{https://github.com/jhudsl/Documentation_and_Usability}{GitHub repository} was created from the \href{https://github.com/jhudsl/OTTR_Template}{OTTR Template repository}. 
Figure~\ref{fig:rendering} illustrates what an OTTR course looks like in development as well as on each platform. 
Edits to the course were done in a text editor. 
When these changes were added to a pull request they were reviewed and eventually made live. Updates were then automatically rendered on GitHub to a \href{https://jhudatascience.org/Documentation_and_Usability}{Bookdown website}. 
This rendering was then used to create a \href{https://www.coursera.org/learn/documentation-usability-cancer-informatics}{Coursera course} and \href{https://leanpub.com/universities/courses/jhu/documentation_and_usability/}{Leanpub course} that are available for certification. 
Course updates have been made since its original publication with OTTR these updates have taken less manual steps. 
Additionally, this course's GitHub repository has received and incorporated updates and new features from the main OTTR repository. 

\section{Discussion}

OTTR is open-source and reduces maintenance burdens.
It allows authors to write content once and publish with Bookdown, Leanpub, and Coursera. 
Menial tasks of course creation are automated by OTTR. 
It is modular and customize-able, allowing authors to choose what features are most useful for their course.
OTTR utilizes the GitHub workflow familiar to many tool developers which encourages collaborative review and transparency. 
This workflow also allows for community contributions.
New OTTR features and changes are offered to course authors through pull requests. 

Thus far, OTTR has reduced our own team's burden for creating courses by allowing us to focus more of our time on content. 
Overall, we believe the creation of OTTR will contribute to greater content creator efficiency, fewer maintenance headaches, and more productive, collaborative, and accessible learning materials in informatics, data science, and beyond.

\subsection{Future directions}

Where feasible, we will continue to add tools, guidance, and documentation to make course creation and maintenance even less of a burden.  

\begin{itemize}
\item \textbf{Increase usability} - We plan to conduct more usability testing to increase the usability of OTTR so that users can create courses with even more ease. 
\item \textbf{Expand features} - We aim to expand features so OTTR is applicable to a greater variety of course contexts and the needs that accompany them.
\item \textbf{Add guidance} - We intend to add guidance in the form of documentation and videos to help users of OTTR overcome any prerequisite knowledge barriers. 
\item \textbf{Build a community} - We hope to build a community of course creators who can support each other's education endeavors with OTTR tools. 
\end{itemize}

OTTR is a constantly developing set of tools since it is being used and developed by our team and others as we create more course content. 
We greatly value feedback from users of OTTR about how it can be improved. 
Ideas and feedback can be provided in the form of \href{https://github.com/jhudsl/OTTR_Template/issues/new/choose}{GitHub issues}. 
There's also a \href{https://github.com/jhudsl/OTTR_Template/wiki/Getting-Help-(Google-Group)}{Google Group} that we encourage users of OTTR to join. 

\section*{Acknowledgements}

We thank Ira Gooding for his invaluable advice regarding the Leanpub and Coursera platforms. 

\section*{Declaration of interest statement}

The authors declare no competing interests. Individual courses we have created on Coursera and Leanpub do generate course revenue, but we do not obtain revenue for any courses other individuals create using OTTR. 

\bigskip
\begin{center}
{\large\bf SUPPLEMENTARY MATERIAL}
\end{center}

\begin{description}

\item \href{https://cran.r-project.org/web/packages/ottrpal/index.html}{ottrpal is available on CRAN} 

\item Existing course creation tools: A table that compares the workflows and features of Software Carpentries, Galaxy Training, and OTTR (JPG table file)

\end{description}

\begin{figure}
\includegraphics[width=1\linewidth]{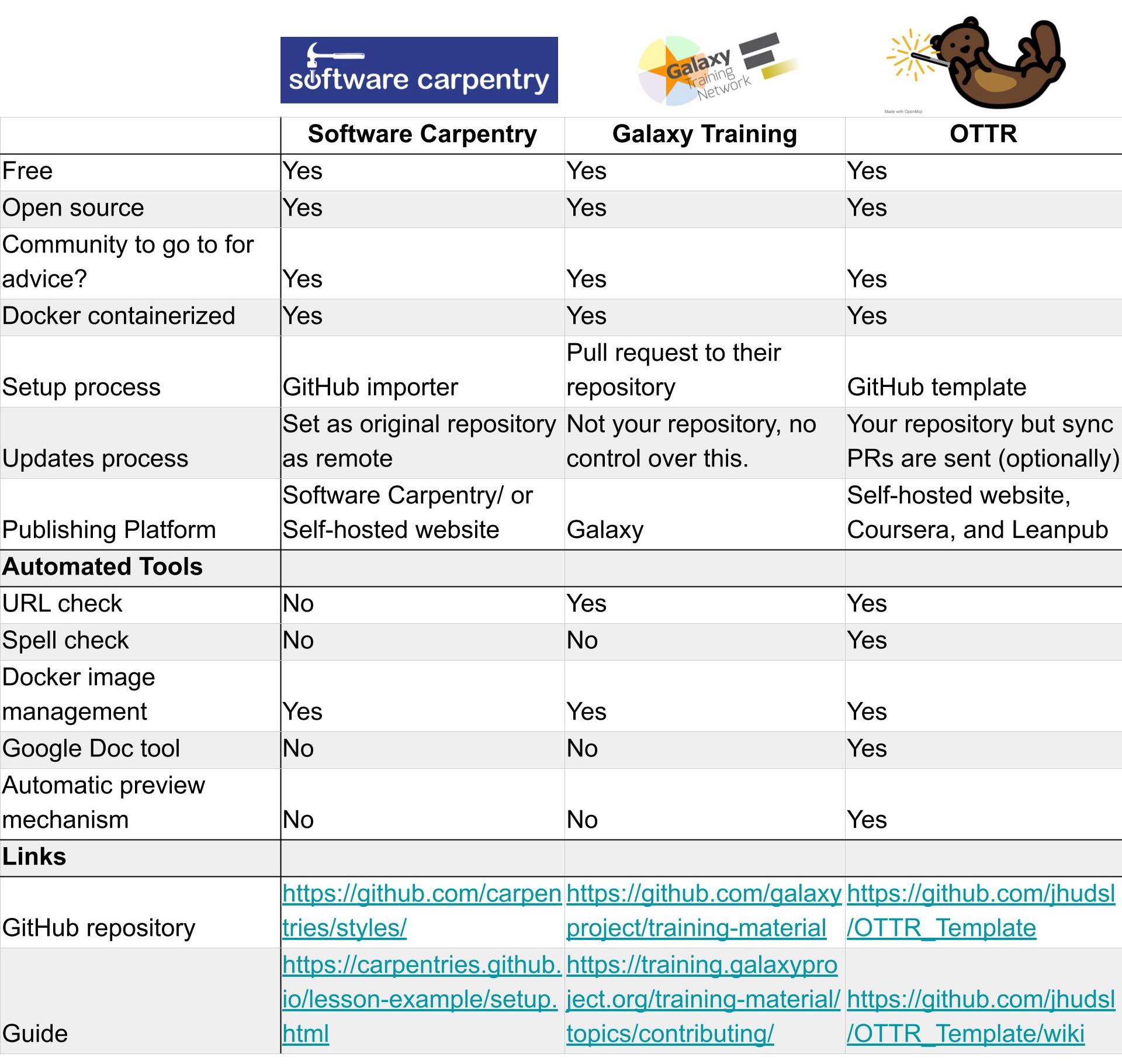}
\label{fig:compareplatforms}
\end{figure}

\section*{Additional Information}

\subsection*{Funding}

This work was supported by the National Cancer Institute under Grant UE5CA254170 and the National Human Genome Research Institute under Grant U24HG010263.

\bibliography{references}

\end{document}